# Interlayer Coupling in Two-Dimensional Semiconductor Materials

Zhiming Shi[1,†], Xinjiang Wang[1,†], Yuanhui Sun[1], Yawen Li[1], and Lijun Zhang[1,2,*]

[1]*Key Laboratory of Automobile Materials of MOE and College of Materials Science, Jilin University, Changchun 130012, China*

[2]*State Key Laboratory of Superhard Materials, Jilin University, Changchun 130012, China*

[†]*These authors contributed equally*

[*]*Address correspondence to: lijun_zhang@jlu.edu.cn*

**ABSTRACT**

Two-dimensional (2D) graphene-like layered semiconductors provide a new platform for materials research because of their unique mechanical, electronic and optical attributes. Their in-plane covalent bonding and dangling-bond-free surface allow them to assemble various van der Waals heterostructures (vdWHSs) with sharply atomic interfaces that are not limited by lattice matching and material compatibility. Interlayer coupling, as a ubiquitous phenomenon residing among 2D materials (2DMs) systems, controls a thin layer exfoliation process and the assembly of vdWHSs and behaves with a unique degree of freedom for engineering the properties of 2DMs. Interlayer coupling provides an opportunity to observe new physics and provides a novel strategy to modulate the electronic and optoelectronic properties of materials for practical device applications. We herein review recent progress in the exploration of interlayer coupling in 2D semiconducting vdWHSs for potential applications in electronics and optoelectronics.

# 1 Introduction

Stimulated by the discovery of promising properties of graphene [1-6], two-dimensional materials (2DMs) with unique physical and chemical properties have become one of the most popular topics in material science. 2DMs have attracted great attention because they exhibit excellent potential for applications in photovoltaic and optoelectronic devices, field-effect transistors, electrocatalysts, topological insulators, etc. [7-19]. The research of 2DMs started with graphene following the pioneering article by Novoselov and Geim [1], and it was demonstrated very soon after that other 2DMs also exhibit amazing properties [20]. Layered hexagonal boron nitride (*h*-BN) is predicted theoretically to increase the band gap in graphene when graphene was deposited on *h*-BN [21]. This stimulated a significant increase in research on *h*-BN and led to an understanding that *h*-BN can be used as an ideal substrate for graphene electronic devices [22]. Very soon after the research of graphene and *h*-BN was conducted, investigations on many other 2D semiconductors demonstrated that the electronic properties of the layered materials family undergo great changes as they approach monolayer dimensions [23, 24]. Therefore, 2DMs, including graphene, *h*-BN, transition metal dichalcogenides (TMDCs), monoelemental 2D semiconductors (phosphorene, silicene, germanene) [25-29], MXenes [17, 30] and monochalcogenides (GaSe, etc.) [31, 32] are attractive to many researchers.

In moving from bulk to monolayer dimensions, many interesting properties are discovered, implying that interlayer coupling plays an important role affecting both the mechanical and electrical properties of the materials [33-35], as shown in Fig 1a. Graphene shows a range of unusual properties when compared to graphite, such as record stiffness and extraordinary electronic transport properties [1, 5, 36]. The same can also be found for other 2DMs. For instance, black phosphorus and TMDCs ($MX_2$, M = Mo, W; X = S, Se and Te) undergo indirect to direct band gap transitions as the thickness is reduced from bulk to monolayer thickness [8, 28]. These findings provide a new dimension to modify the performance of materials by interlayer coupling engineering. The great in-plane stability provided by strong covalent bonds and relatively weak interlayer van der Waals forces make the van der Waals heterostructures (vdWHSs) exhibit excellent horizontal stiffness and make flexible perpendicular assembly possible. Naturally, stacking these 2D semiconductors layer-by-layer in any desired sequence to build hetero architectures with entirely new functions provides novel insight into materials or device design [37-41]. Geim vividly described this process as building [39] with Lego bricks. Recently, the possibility of realizing multilayer vdWHSs has been experimentally confirmed [42-46]. In this review, we will discuss the recent progress of interlayer coupling and applications in two-dimensional semiconductors. First, we will briefly introduce the distinctive properties induced by interlayer coupling for both homo and hetero 2DMs. Second, we will review the experimental and theoretical approaches for detecting inlayer coupling. Third, we will discuss the novel physical properties generated by the interlayer coupling and explain their basis. Fourth, a brief introduction of the application for these findings will be given. At last, we will present our perspective on this field based on our viewpoint. We expect that this review will be helpful to extend the applications of 2DMs with desirable performance, such as optoelectronic devices, field-effect transistors (FETs), light-emitting devices (LEDs), photodetectors or catalysts.

## 2 Different Types of Interlayer Coupling

### 2.1 Interlayer Coupling in Well-Stacked Multilayer 2DMs

The simplest prototype is well-stacked multilayer 2DMs. By analyzing the vibration properties of monolayer and multilayer structures, the effects induced by interlayer coupling can be easily detected [47-50]. Graphene is still a pioneer material of interest. Bilayer graphene consists of two monolayers of graphene stacked in the Bernal type (AA or AB, as shown in Fig 1b and c), making it distinct from monolayer graphene. The behaviors of bands near the Fermi level are different. The low-energy band dispersion of intrinsic bilayer graphene is quadratic (an linear in monolayer graphene) with massive chiral quasiparticles [51, 52] rather than massless ones. Furthermore, stacking two functional graphene layers (doping or gating) with entirely different properties together achieved a band gap as high as 300 meV [51, 53-57]. Bilayer or multilayer graphene may also provide increased electrical, thermal, strength [58, 59], or optical properties [52, 60].

TMDCs are the next most popular 2D semiconductor family after graphene. The most interesting feature of TMDCs is their indirect to direct band gap transition induced by thinning layers from bulk to monolayer [23, 61-65]. Previous studies have suggested that strong electronic interactions between the two individual monolayers in bilayer TMDCs are in response to the direct to indirect gap transition [66-68]. Meanwhile, different stacking configurations of bilayer TMDCs also displayed various optical absorption spectra, band structures [69] and elastic properties. For example, the in-plane stiffness values of monolayer and bilayer $MoS_2$ are quite different, e.g., 180±60 Nm$^{-1}$ (effective Young's modulus of 270±100 GPa) and 260±70 Nm$^{-1}$ (200±60 GPa), respectively.

Recently, multilayer and monolayer (phosphorene) black phosphorus has been realized by mechanical exfoliation [28, 29]. The structural, electronic, and optical properties of multilayer phosphorene have been extensively studied both experimentally and theoretically [28, 29, 70-74]; however, to the best of our knowledge, the role of interlayer coupling remains debatable or unclear. For instance, whether the band-edge transition is induced by interlayer coupling is still controversial [75-79]. Cai et al. attempted to employ quantum confinement effects to interpret the thickness dependence of the band gap [80]. Tran et al. predicted that the band gap-thickness dependence decay behavior of phosphorene is significantly different from normal quantum confinement results [72], which may demonstrate that the quantum confinement effect is not the only reason attributed to the evolution of the band gap as a function of the number of layers [75-79]. Another observation of the faster up-shift of the valence band maximum (VBM) than of the conduction band minimum (CBM) is also opened [81].

## 2.2 Interlayer Coupling in Twisted Multilayer 2DMs

Weak van der Waals interactions lead to individual monolayers having the same orientations, which is different from naturally existing layer orientations, and these results have been experimentally confirmed by probing different vibrational modes by Raman spectroscopy [47, 49, 82-84]. Subsequently, interlayer coupling can also be modified by changing the twist angles in multilayer 2DMs due to destruction of the symmetry. In twisted bilayer graphene, two Dirac semimetal monolayers form large Moiré patterns, and the Dirac-like linear dispersion remained but with a lower Fermi velocity than monolayer graphene [85-87]. Furthermore, the energy of the van Hove singularities (VHSs) induced by the crossing bands depends on the twist angle [88].

TMDCs also show twist angle dependence properties [89-91]. Liu et al. and Huang et al. have revealed the evolution of interlayer coupling in twisted bilayer $MoS_2$ [89, 91]. Theoretical simulations demonstrated that evolution originates from different interlayer separations due to different interlayer coupling interactions between the two $MoS_2$ layers in different stacking configurations [89]. Zande et al. confirmed that the indirect optical transition energy and second-harmonic generation are sensitive to the twist angles, while direct optical transition energies and Raman vibration modes are insensitive [90].

## 2.3 Interlayer Coupling in van der Waals Heterostructures (vdWHSs)

Assembly of 2D semiconductors into vdWHSs (Fig 1d) provides a promising approach to obtain desirable properties. The pioneering effort involves stacking graphene onto *h*-BN to form vdWHSs [22]. The smooth surface and chemical inertness of BN show BN's high potential as an ideal candidate to weaken the substrate effects. Further attempts have been made to encapsulate graphene with another layer of *h*-BN on top to protect graphene from the environment, such as active gas molecules, and results in greatly enhanced carrier mobility (higher than 100,000 $cm^2\ V^{-1}\ s^{-1}$) [92]. Similar strategies can also transfer to TMDC systems and lead to improved performance [93-95]. Encapsulation is described as a practical method to stabilize unstable 2DMs, such as black phosphorene [96] and organic-inorganic hybrid perovskites [97], in which device encapsulation exhibits a performance as good as that of the exposed system under both vacuum and ambient conditions. Additionally, BN is also an ideal dielectric layer because of its large band gap [98], which cannot be tunneled despite its thickness of one atom. Lattice mismatch and orientation misalignment cause Moiré patterns in vdWHSs [99]. Moiré patterns create new superlattices in electronic structures, with new Dirac points dependent on the wavelength of the Moiré pattern; the Fermi velocity is significantly decreased, leading to anisotropic transport within the graphene layer [100]. Theoretically, large strain induced by lattice mismatch between graphene and *h*-BN could increase the band gap in graphene; however, this has not yet been experimentally confirmed [101].

## 3 Experimental Observation and Theoretical Prediction of Interlayer Coupling

### 3.1 Photoluminescence (PL) Spectra

Because of the zero band gap of graphene, the PL method is difficult to use in graphene systems. Meanwhile, indirect to direct band gap transition results in significant enhancement in PL signal (as shown in Fig 2a for mono and bilayer $MoS_2$) [23, 61, 102]. As illustrated in Fig 2b and c, the peak moves to higher energies for mechanically exfoliated $MoS_2$ and $MoTe_2$ samples as the layer number decreases due to the stronger quantum confinement effect [23, 102]. Peaks A and I originate from the direct and indirect band gaps, respectively. Peak I vanishes as the sample thins to a monolayer thickness since the material transitions to direct band gap. Peak B forms due to the combined effect of interlayer coupling and spin-orbit coupling. These trends are consistent with the relationship between band-edge evolution and layer number. Similar results are also observed in $WS_2$ and $WSe_2$ (Fig 2d-g), where the emission intensity of monolayers is 100–1000 times stronger than that of bulk materials [103]. At the same time, the difference between the bulk and monolayer $MoS_2$ samples is more modest than that between bulk and monolayer $WS_2$ because of the polarization of different chemical bonds between S-W-S and S-Mo-S, which is also reported in Ref [104]. A specific PL temperature dependence behavior is found in $MoSe_2$ [105]. In monolayer $MoSe_2$, the PL intensity decreases with temperature but unexpectedly increases for a few-layer sample. The nearly degenerate indirect and direct gaps in few-layer $MoSe_2$ samples should be attributed to the unusual temperature dependence. The interlayer spacing will thermally increase from the equilibrium position at 0 K. Consequently, the interlayer coupling is weakened, which results in indirect-direct band gap transitions. In contrast, a similar process is difficult to induce because the indirect and direct band gaps are well-separated in bilayer $MoS_2$, and therefore, the band gap transition cannot be thermally approached unless the two individual layers are separated from each other. For $ReS_2$, the PL intensity abnormally decreases [106] from the bulk to a monolayer. Density functional theory (DFT) calculations showed direct band gaps for both the bulk and monolayer $ReS_2$. Furthermore, the band structures are not sensitive to stacking and thickness, which is much different from most other TMDCs due to a relative weak interlayer coupling strength.

The change in the interlayer coupling properties induced by twist angles between the two TMDCs layers also effects PL emission. Twisted bilayer $MoS_2$ arranged at an arbitrary angle is a promising model for studying the variation in PL emission with twist angle [89-91, 107]. We subsequently summarize the main findings. In Ref [91], a twist angle dependence PL of bilayer MoS$_2$ was reported. The systematic results of the investigation are displayed in Fig 3. As we can see, the intensity ratio exhibits a periodic oscillatory behavior as a function of the twist angle. The maximum intensity ratio peaks appear at θ = 0°, 60°, and 120°, and the minimum intensity peaks appear at 30° and 90° with respect to the D$_{3h}$ symmetry of monolayer MoS$_2$. The interlayer distances, trion binding energies, and E$_B$ - E$_A$ show similar behaviors, suggesting that interlayer couplings at 30° and 90° are totally different from those at 0° and 60°. Both of the experimental observations concluded that at a twist angle of 0° or 60°, the interlayer coupling is strongest mainly due to the small separation distances, and variation of the A$^-$/A PL intensity ratio is mainly attributed to the difference of trion binding energy with different twist angles. Similar results are also reported by Zande and Liu [89, 90]. PL spectra of few-layer MoS$_2$ with different stacking configurations is reported in Ref [108]. A stronger PL response is found in pyramidal MoS$_2$ flakes [109] because the stacking is different from 2H.

### 3.2 Raman Spectroscopy

Raman spectroscopy is a non-destructive characterization tool used to detect the materials features, such as phonons, electron–phonon coupling, band structures and interlayer coupling in 2DMs [110-112]. The interlayer coupling results in a significant change of interlayer force constants. In monolayer graphene, there are six normal modes, $\Gamma = A_{2u} + B_{2g} + E_{1u} + E_{2g}$, which are Davydov doublets in graphite, written as $\Gamma = 2(A_{2u} + B_{2g} + E_{1u} + E_{2g})$. The $E_{2g}$ mode relates to interlayer shear modes, involving relative motion between adjacent layers. The cases in TMDCs are more complicated. For 2H bulk TMDCs with a D$_{6h}$ point group, the 18 normal vibration modes at the Γ point are $\Gamma = A_{1g} + 2A_{2u} + 2B_{2g} + B_{1u} + E_{1g} + 2E_{1u} + 2E_{2g} + E_{2u}$, where $A_{1g}$, $E_{1g}$ and $E_{2g}$, are Raman (R) active modes [50, 110, 113]. Tan et al. successfully employed the linear chain and improved the model to systematically analyze the shear mode of few-layer 2D homo- or heterostructures [48, 82, 83, 95, 114, 115] and suggested that the corresponding Raman peak can be used to measure interlayer coupling not only for Bernal stacking but also for twisted systems. The theoretical results are perfectly reproduced in the experiments. Fig 4a shows representative Raman spectra for single- and multiple-layer MoS$_2$ [116]. Only two Raman-active modes ($E_{2g}^1$ and $A_{1g}$) among four of the 2H bulk MoS$_2$ are clearly observed at approximately 400 cm$^{-1}$. However, the two other Raman-active modes ($E_u$ and $A_{2u}$) could not be detected due to selection rules. Importantly, $E_{2g}^1$ redshifted, while the $A_{1g}$ vibration blueshifted, with increasing sample thickness. When there are more than four layers, both of the mode frequencies converge to bulk materials. Low-frequency Raman modes are also studied in Ref [117] and [118]. The $E_{2g}^2$ mode undergoes a blueshift from 22 cm$^{-1}$ for bilayer to 32 cm$^{-1}$ for bulk with increasing thickness and is identified as the interlayer shear mode [84, 119]. Evolution of the polarized low-frequency Raman spectra as a function of the number of layers is shown in Fig 4b for $MoS_2$ and $WSe_2$ [119]. In both $MoS_2$ and $WSe_2$, the S1 ($E_{2g}^2$ mode) peaks gradually shift to low-frequency as the thickness from bulk (32 cm$^{-1}$) to bilayer (22 cm$^{-1}$) decreases. DFT calculations demonstrate that the S1 peak comes from the highest frequency shear mode. Moreover, the B1 peak contributes to the lowest frequency interlayer breathing mode. This assignment agrees with the B1 peak disappearance in the $\bar{z}(xy)z$ polarization configuration. Therefore, low-frequency modes are used as an efficient method to detect not only the number of layers but also different stacking configurations (e.g., 2H or 3R) [120]. Additionally, high-frequency intralayer vibration modes in multilayer 2DMs have also affected the interlayer coupling [113, 115, 121]. Few-layer $MoTe_2$ has been selected to investigate the Davydov splitting of high-frequency modes [115]. The van der Waals model, in which only adjacent interlayer coupling is considered, is used to understand the evolution of Raman spectra with respect to layer number. The observation of the $A_1'$ ($A_{1g}^2$) modes of Davydov splitting are attributed to the resonance frequency of the Raman intensity by 1.96 eV excitation, whose energy is close to the energy of the B' exciton in few-layer $MoTe_2$. The anisotropy of 2DMs also affects the interlayer vibration modes. Tan et al. reported $ReS_2$ with two stable stacking orders, isotropic and anisotropic [122]. Two interlayer shear modes appear in anisotropic stacking, while just one shear mode is observed in isotropic stacking. This frequency distinction identifies unexpected strong interlayer coupling.

Raman scattering also plays an important role in the detection of vdW interactions in TMDC vdWHSs [124, 125]. Ref [124] is a comprehensive study of $MoS_2$-graphene, $MoS_2$-mica and $MoS_2$-$WS_2$ nanostructure and is supplemented with a $MoS_2$-$WSe_2$ study [125]. When stacking $MoS_2$ and graphene together, the Raman modes changed slightly, which indicated that interlayer coupling is negligible between graphene and $MoS_2$. For the vdWHSs composed of n-layer (n = 1, 2) graphene and m-layer (m = 1, 2) $MoS_2$ on Si/SiO$_2$, the graphene peak is located at 1583–1584 cm$^{-1}$ in all cases. Additionally, the second-order Raman response of free-standing graphene shows a blueshift from 2685 cm$^{-1}$ to 2699 cm$^{-1}$ after being transferred to $MoS_2$. In $MoS_2$, both the $E$ and $A$ modes shifted when the monolayer was sandwiched with graphene. In particular, the $E_{2g}^1$ peak slightly shifted from 385 to 384 cm$^{-1}$ when capped with one more layer of graphene, and the $A_{1g}$ mode shifted from 403 to 405 cm$^{-1}$. These peaks will not change further when a second layer of graphene is stacked on top. This demonstrates that $E_{2g}^1$ and $A_{1g}$ shifts are induced only by the nearest direct contact with graphene and not by the thickness of the graphene layers on $MoS_2$. Similarly, for bilayer $MoS_2$, the $A_{1g}$ peak moves from 405 to 406.5 cm$^{-1}$ when encapsulated with one or two layers of graphene. The changes in the $E_{2g}^1$ and $A_{1g}$ positions exhibited after encapsulating $MoS_2$ are similar to those of pristine $MoS_2$ when a monolayer increases in thickness to few-layer $MoS_2$ [116, 126]. Similar analyses are also used to investigate other vdWHSs, such as $MoS_2$-mica and $MoS_2$-$WS_2$ [124]. Another effective probe in vdWHSs is interlayer breathing modes [123]. Raman spectra of several bilayer vdWHSs are displayed in Fig 4c. A peak at approximately 32 cm$^{-1}$, which is not observed in bilayer homostructures, emerges for both $MoS_2$/$WSe_2$ and $MoSe_2$/$MoS_2$ vdWHSs [123]. This implies that the peak comes from interlayer coupling between two different adjacent TMDC layers. Because the interaction of Se and S atoms replaced the interaction of S and S atoms (in homostructures) to contribute the interlayer coupling in $MoSe_2$/$MoS_2$, the interlayer coupling in the different vdWHSs is totally different from that in the homostructures. The twist angle-dependent low-frequency modes demonstrate that interlayer coupling is not only sensitive to the atomic species but also to the relative orientation of the stacked layers.

### 3.3 Angle-resolved Photoemission Spectroscopy (ARPES)

Angle-resolved photoemission spectroscopy (ARPES) can directly measure the band structures of materials. Therefore, it is a powerful tool to detect differences in band structures induced by interlayer coupling. Many achievements have been published. Taisuke et al. observed two cones and VHSs in twisted bilayer graphene (twist angles > 5°) by ARPES measurements, as shown in Fig 5a-c. A small band gap opens at the Brillouin zone boundaries due to a periodic potential induced by the Moiré interlayer coupling [127]. Jin et al. revealed the dependence of the band structures on thickness in layered $MoS_2$ [128] (Fig 5d), and the results agree with the theoretical predictions. Zhang and co-workers report the direct observation of an indirect-direct band gap transition (Fig 5e) induced by interlayer coupling by using ARPES on high-quality $MoSe_2$ with different thicknesses [129]. They found that the experimental band gap is larger than the theoretically predicted band gap, and a significant spin-splitting as great as 180 meV was obtained at the VBM of monolayer $MoSe_2$.

### 3.4 Theoretical Confirmation of Interlayer Coupling

Theoretical calculations are helpful to understand and predict experimental observations and can be employed to investigate interlayer coupling. They provide a convenient approach to study the effect of interlayer distance, stacking configuration and twist angles, and external electric field among other parameters on interlayer coupling. For instance, the band structure of bilayer graphene can be tuned by the interlayer distance and external electric field (Fig 6a and b) [130]. The band gap linearly increases with the vertical electric field when the interlayer distances are smaller than 0.4 nm. Deniz and his colleagues selected trilayer black phosphorene as a prototype to reveal the dependence of band structures on the stacking configuration. The band gaps are 0.80, 0.76 and 0.49 eV for ABA, AAB and ACA, respectively [131]. Because the overlap of the $p_z$ orbitals dominate the band gaps in different stacking configurations, the vibrational frequencies can be calculated for phonon dispersion analysis and Raman spectroscopy prediction, as shown in Fig 6c and d [132]. In MoS₂, Raman-active modes $A_{1g}$ (out-of-plane) and $E_{2g}^1$ (in-plane) blueshift and redshift, respectively, with increasing layer numbers. The layer dependence trend is attributed to the surface effect since larger force constants at the surface of the thin film can make a difference [133]. Moreover, the perpendicular elastic modulus of graphene and graphene oxide vdWHSs has also been investigated by DFT calculations [134], which showed that the interlayer elasticity is sensitive to intercalated molecules. In particular, the interlayer elastic modulus decreased and then increased with the intercalation of H₂O molecules between the graphene oxide layers. Jun and co-workers [135] predicted the charge redistribution of MoS₂/MoSe₂ bilayer vdWHSs induced by the Moiré pattern. The VBM and CBM of the vdWHSs are affected by the twist interlayer coupling periodic potential (Fig 6e). The CBM localization is attributed to competition between the vertical dipole moment and the lateral potential difference at different regions. In addition to the modification of properties, the strength of the interlayer coupling directly influences the binding energy among layers[136]. As shown in Fig 6f, the exfoliation/binding energy increases with thickness due to the presence of longer ranges of interlayer interactions, and the energy converges for all considered materials when the number of layers is more than 4.

## 4 Novel Properties and Applications Induced by Interlayer Coupling

### 4.1 New Physics Generated by Interlayer Coupling

#### 4.1.1 Band Gap Opening in Graphene

The most important obstacle is the absence of an energy gap separating the valence and conduction bands of graphene, which is described as a semimetal. As a consequence, electrical conduction cannot be switched off using control voltages, which is an essential feature for the operation of conventional transistors. Theoretical physicists have first provided a new method of stacking another monolayer of graphene to construct bilayer graphene [51], and the band gap can be tuned by applying an electric field (Fig 7a) [55]. Electronic wave-function asymmetry induced by interlayer coupling and transverse electric field introduces a band gap in bilayer graphene, which is not observed in monolayer graphene. The prediction has been experimentally confirmed by Zhang et al. (Fig 7b) [137]. A gate-tunable band gap as large as 250 meV, which is higher than the room temperature thermal energy (25 meV), can be achieved, emphasizing the intrinsic potential of bilayer graphene for nanoelectronics. Furthermore, Taisuke Ohta also realized band gaps in bilayer graphene by controlling carrier density [54].

As we mentioned above, the principle of band gap opening is breaking the symmetry of graphene. Therefore, the Moiré pattern in bilayer structures formed by twist angles or lattice mismatch should be expected as an ideal candidate for breaking the symmetry. Indeed, both theoretical and experimental studies have confirmed this assumption in twisted graphene bilayer [139] or graphene/*h*-BN vdWHSs [140-142]. Fan et al. theoretically revealed that the band gaps are very sensitive to interlayer distance [138]. When the interlayer spacing is larger than 4 Å, the band structure is almost the same as that in graphene (Fig 7f). With decreasing interlayer spacing, stronger interlayer coupling can be observed by measuring the charge redistribution (Fig 7g). The equivalence of the two sublattices in graphene is destroyed, resulting in a small gap. Obviously, the stronger the interlayer coupling is, the larger the band gap will be. Moreover, the high mobility of graphene remains in the graphene/*h*-BN bilayer. Similar results are also obtained in other graphene vdWHSs, such as graphene/g-$C_3N_4$ [143] and graphene/GaN [144].

### 4.1.2 Role of Interlayer Coupling on the Evolution of Band Edges

Bulk TMDCs show an indirect gap. At the same time, when thinned to monolayer TMDCs, the band gap in VIB group TMDCs increases, and more importantly, the materials become direct gap semiconductors [23, 61-63, 145-147]. This is responsible for the experimentally observed significant enhancement of the PL intensity [23, 61, 102]. The evolution of the band edges with increasing layer number is illustrated in Fig 8a-c [146]. DFT calculations indicate an indirect band gap in bulk $MoS_2$ located between $\Gamma_V$ and $T_C$ (left in Fig 8a) of the Brillouin zone and a direct band gap in monolayer $MoS_2$ at *K* point (right in Fig 8a). The interlayer electronic wave-function overlap induced by interlayer coupling governs the indirect to direct band gap transition, as shown in Fig 8b. $K_V$ and $K_C$ almost originate from the intralayer located states, which is the dominant factor in monolayer $MoS_2$ when interlayer coupling is absent. In the bulk phase, the delocalized $\Gamma_V$ and $T_C$ states are affected more by interlayer coupling than localized $K_V$ and $K_C$ and shift up and down in energy to become the band edges. The evolution of band gaps in $MoS_2$ based on the number of layers $n = 1$ to 10 is shown in Fig 8c. The interlayer delocalized $\Gamma_V$ and $T_C$ states show that the lower boundary of the energy level groups increases in energy, and the higher boundary decreases, almost converging when $n > 4$. Furthermore, the intralayer localized $K_V$ and $K_C$ states display a pretty narrow energy span. Wang et al. [148] reported the evolution of the band gap of few-layer phosphorene and revealed that interlayer coupling played a dominant role in decreasing both the band gap and the carrier effective mass as the thickness increased. Monolayer phosphorene showed an indirect band gap and transitioned to a direct band gap with additional stacking due to strong interlayer coupling on the $p_z$ orbital, as shown in Fig 8d. Recent study of layer-dependent variation of band gap and band-edge energy levels in indium selenide by some of the authors [149] indicated that in addition to the proposed quantum confinement effect [150], the strong interlayer coupling plays an important role in determining layerdependent electronic and phonon properties.

### 4.1.3 Wide-Range Tunability of Black Phosphorene Band Gap

An important effect in the black phosphorus system is the band gap dependence on thickness. [28, 29, 70, 72, 151-153]. The band gaps can be tuned from approximately 0.3 (bulk) to 1.0 (monolayer) eV by controlling the layer thickness (Fig 9b). The electronic band structure of black phosphorus with different layers exhibited direct band gaps at the Γ point of the Brillouin zone, as shown in Fig 9a [151]. We can observe a redshift in the band gap by adding layers. In contrast, unlike the band gaps, band-edge dispersions are almost unaffected by interlayer interactions. Similar results have also been confirmed by DFT calculations at different levels (inset of Fig 9b) [152]. In Fig 9c, Qiao and co-workers employed real-space spatial distribution of wavefunctions for edge states to reveal the interlayer coupling between two layers. Clear interlayer hybridization of electronic states is visible for VB1 (in red rectangles), implying chemical bonding features. Meanwhile, the VB2, CB1 and CB2 states exhibited anti-bonding features. The wavefunction overlap rather than weak van der Waals contributed most to the interlayer coupling for black phosphorene. Strong electron-phonon coupling due to interlayer coupling introduced a band gap reduction of 0.5 eV from monolayer to bilayer black phosphorus. More stacking layers resulted in stronger interlayer coupling, and therefore, the band gaps continued to decrease with the addition of more layers and reached 0.3 for bulk structures. This effect can also be used to interpret the change in the lattice parameters from monolayer to bilayer structures.

### 4.1.4 Mobility Engineering

As we discussed before, interlayer coupling will affect both the phonon and electric properties of the materials; meanwhile, the two factors have very important roles in the carrier mobility of devices. Therefore, many contributions are donated to improve the carrier mobility of devices by tuning interlayer coupling [28, 93, 94, 152, 154-156]. Wang et al. attempted to encapsulate the graphene with *h*-BN [156] to construct 1D metallization contact, and the device exhibited ultrahigh room temperature mobility comparable to the theoretical limit. Similar strategy has also been applied in $MoS_2$ field-effect transistors (FETs) [93, 94]. Liu et al. reported the record-high field-effect mobility up to 1300 $cm^2 V^{-1} s^{-1}$ in $MoS_2$ at low temperature [93]. As shown in Fig 11, BN encapsulation can effectively increase the carrier mobility by more than 200%. Due to interlayer coupling, the BN layer can reduce phonon scattering and screen electric scattering. In contrast, the case in black phosphorus is completely different. The interlayer and stacking-induced in-plane overlap of the VB wavefunctions increases with increasing layer number (in Fig 10c), but this effect is absent in the monolayer material [152]. Therefore, the hole-dominated mobility of few-layer black phosphorus increased ten times (from 640 to 6400 $cm^{-1} V^{-1} s^{-1}$) when the layer number was increased from 1-5 layers.

### 4.1.5 Interlayer Charge Transfer

Interlayer charge transfer/redistribution induced by interlayer coupling is an important and useful phenomenon in vdWHSs, particular in TMDCs. Hong and co-workers observed ultrafast interlayer hole transfer between $MoS_2$ and $WSe_2$ vdWHSs, approximately 50 fs after optical excitation [157]. Rivera et al. found that the carrier lifetimes of interlayer excitation in the $MoSe_2$–$WSe_2$ systems were as long as 1.8 ns, which is much longer than those in intralayer excitations [158], as shown in Fig 11a-d. The photoluminescent emission energy and intensity can be tuned by applying an external electric field from the gate. Matthew et al. observed additional photoluminescence (Fig 11e) peaks induced by the recombination of trions in coupled $WS_2$–$MoSe_2$ vdWHSs. Trions formed via electrons in the $MoSe_2$ layers and excitons in $WS_2$ layers [159]. According to theoretical predictions, most of these experimental observations can be attributed to the formation of a type II band alignment in TMDC semiconductors, which improves electron–hole separation for light harvesting [160-163], as shown in Fig 11f. In type II band alignment, the VBM and CBM reside in different layers, resulting in spatially separated electrons and holes, and recombination processes are blocked. Therefore, the exciton lifetime is longer than that in the monolayer material.

**4.1.6 Secondary Dirac Points (SDPs), Landau Level Renormalization, Hofstadter's Butterfly**

Another effect is the emergence of secondary Dirac cones (SDCs) in graphene system. From the lower panel in Fig 12a, in addition to the first Dirac cones (FDCs) appearing at the center of the superlattice Brillouin zones (SBZs) [164, 165], six satellitic SDCs also exist at the edges of the SBZ for both the conduction and valance bands of graphene [142]. The presence of SDCs has been certified by experimental observations and theoretical simulations [100, 142, 164, 166-168]. The observed SDCs in experimental dI/dV curves for different Moiré wavelengths (red and black arrows in Fig 12d) and the energy difference between the SDCs and the FDC diminish with increased wavelength (Fig 12e) [100, 169]. It is also difficult to observe SDCs in highly twisted structures. In addition to the locations, ARPES measurements show that the gap at SDCs is approximately 100 meV, which is similar to the magnitude at FDC (Fig 12b and c) [164]. The Moiré potential induced by the interlayer coupling effect leads to the emergence of Hofstadter butterfly states under an applied magnetic field [142, 170-172]. Under a periodic potential, the discrete and degenerate Landau levels from graphene are split into the Hofstadter spectrum; the "Hofstadter minigaps" are separated by a hierarchy of self-similar minigaps [173]. Magnetoresistance data indeed show strong effects from the superlattice (Fig 12f) [142]. Recently, the demonstrated intersections between the central and satellite fans occur at $\varphi = \varphi_0/q$ (Fig 12g), where $\varphi$ is the magnetic flux per superlattice unit cell, and q is a positive integer [142, 170, 171]. The observation of the Hofstadter butterfly is most obvious in the regime $\varphi/\varphi_0 > 1$ and is almost impossible in the monolayer graphene system.

**4.1.7 Plasmon Properties**

The plasmon–phonon polaritons induced by interlayer coupling in graphene/h-BN vdWHSs are shown in Fig 13 and Fig 14. The result of electromagnetic coupling of graphene surface plasmon polaritons ($SP^2$) [174, 175] and phonon polaritons in h-BN [176, 177] was categorized as surface plasmon–phonon polaritons ($SP^3$) and hyperbolic plasmon–phonon polaritons ($HP^3$) based on the propagation range [178, 179]. The propagation range of $SP^3$ was localized at the graphene/h-BN surface, while $HP^3$ propagated throughout the entire vdWHS. Coupling was induced by sufficient interaction between the lattice displacement in graphene/h-BN vdWHS and free carriers in graphene [180]. As shown in Fig 13c, d [181, 182], new $SP^3$ modes arise at 1365 $cm^{-1}$ and approximately 1370 $cm^{-1}$ compared with free-standing h-BN. Those electromagnetic hybrid materials possessed advantages of both the ability to tune the spectral range and long phonon lifetimes, and the calculated transmission spectra (Fig 14e) [181] indicated that the intensity of new $SP^3$ modes is sensitive to the distance between graphene and h-BN. In addition, the degree of plasmon-phonon coupling can be tuned by changing the number of h-BN layers (Fig 14b,c) [183]. Thus, electromagnetic coupling can be applied for sensitive detection and can act as a plasmon filter to sort out specific propagation modes. In addition, tunable electromagnetic coupling provides new research directions for tailored optical properties.

**4.2 Applications of van der Waals Stacking in vdWHSs**

**4.2.1 Optoelectronic Applications**

The high carrier mobility of graphene and the good photosensitivity of $MoS_2$ offer the possibility to create efficient phototransistors by constructing graphene/$MoS_2$ vdWHSs [184]. $MoS_2$/$WSe_2$ [155] and $MoSe_2$/$WSe_2$ [158] (with different work functions) combined are used to separate and accumulate electrons and holes in individual layers due to the interlayer charge transfer and high carrier mobility of vdWHS. Because excitation in this vdWHS typically possesses long lifetimes, the recombination energy could be tuned by controlling the interlayer distance. Further functionalization of the different layer with p and n types of doping can realize atomically sharp p-n junction devices [185, 186], and the devices exhibit very high quantum efficiencies. For instance, external quantum efficiencies of GaTe/$MoS_2$ devices are higher than 60% [185]. Furthermore, black phosphorus/$MoS_2$ vdWHSs show tunable performance with an applied external gate voltage [187]. More candidates, such as TMDCs [188] or metal chalcogenides [189], also support various options to create efficient photovoltaic devices. By encapsulating a photosensitive monolayer between graphene electrodes, graphene can adequately extract carriers from the device to form good ohmic contacts with TMDCs.

**4.2.2 Light-emitting Diodes**

Charge carrier separation in the p-n junctions leads to light emission when electron-hole recombination occurs [186]. However, p-n junction devices depend on the high-quality synthesis of p- and n-type materials, which have not yet been realized for all 2D semiconductors. Furthermore, the junction resistance is similar to that of p and n electrodes, causing difficulty in tuning the current distribution. A straightforward strategy to design the device is injecting carriers using highly conductive transparent electrodes. With this method, the dwell time of the injected carriers should be carefully controlled in the semiconductor crystal because the photoemission process is much slower than the carrier penetration across the junction between electrode and the semiconductor. Withers et al. attempted to control the dwell time by introducing *h*-BN layers to screen the interlayer coupling in order to increase the carriers lifetime inside the TMDC monolayer, thus enhancing radiative recombination [190]. Moreover, the quantum efficiency can be affected by stacking the TMDCs in series [190]. Devices fabricated with $WSe_2$ show higher quantum efficiencies, and the efficiencies improves by increasing the temperature (reaching 20% at room temperature) and injection current [191].

### 4.2.3 Photodetector

Band gaps tunable by varying the thickness offer a flexible method to fabricate photodetection devices. For instance, by tuning the numbers of layers in $MoS_2$, Lee et al. realized wide-range (1.8, 1.65 and 1.3 eV for mono-, bi- and tri-layered, respectively) photodetection devices [192], from ultraviolet to infrared wavelengths [193]. These devices exhibited a photoresponsivity of approximately 100 mA/W, which is comparable to that of Si-based devices [194]. A similar strategy has also been applied for $WS_2$ [195]. Efficient photocurrent with a high gate tunability was observed in graphene/$MoS_2$/graphene or graphene/$MoS_2$/metal sandwiched structures, and the maximum external and internal quantum efficiency can reach 55% and 85%, respectively [196]. Few-layer $ReS_2$ and black phosphorus are also excellent candidates to fabricate tunable and high efficiency devices [187, 197, 198].

### 4.2.4 Field-Effect Transistors (FETs)

Vertical stacking of 2DMs to fabricate FET devices has also been achieved experimentally [46, 187, 199] due to high mobility of the 2DMs. Yu et al. constructed FETs with $MoS_2$ and graphene [199], and reported an ON/OFF ratio over $10^3$ and a high current density of up to 5,000 A/cm$^2$ at room temperature. In Ref [46], using $WS_2$ instead of $MoS_2$ to build graphene-$WS_2$ vdWHS vertical FET for flexible and transparent electronics showed an enhanced ON/OFF ratio exceeding $10^6$ at room temperature.

### 4.2.5 Catalysis with vdWHSs

Interlayer coupling between two-dimensional materials may improve the catalyst performance. Dai et al. synthesized $MoS_2$ nanosheets on reduced graphene oxide (RGO), which led to a high electrocatalytic activity for the hydrogen evolution reaction (HER) that was better than the activity of bare $MoS_2$ catalysts [200]. They assumed that the enhancement was attributed to more active edge sites on $MoS_2$ and electrical coupling with graphene. Similar cases were found by Shin et al. replacing $MoS_2$ with $WS_2$ [201]. The $WS_2$/RGO heterostructures also acted as excellent catalysts for the HER because conducting RGO sheets assist the fast electron transfer from the electrode to $WS_2$. Yu et al. reported that 2D $MoS_2$/$CoSe_2$ vdWHSs exhibit high HER catalytic activity in acid since strong chemical coupling increases the HER activity of both $MoS_2$ and $CoSe_2$, which is explained by DFT calculations [202].

## 5. Concluding Remark

In recent years, there has been a rapid increase in interest and tremendous progress in the field of interlayer coupling between 2D semiconductors. The lack of dangling bonds on the surface of 2D semiconductors ensures stable chemical and physical properties, and the existing interlayer coupling allows considerable flexible fabrication of completely different monolayers by restacking without traditional constraints such as lattice mismatch. Atomically sharp interfaces and distinguishable electronic properties provide a novel platform to study fundamental physics and promising applications. However, challenges still exist. Scalable fabrication of monolayer samples and vdWHSs remains a primary barrier for future applications. Many efforts have been devoted in developing synthesis strategies, but progress is currently not sufficient. For example, the physical and chemical properties of chemically synthesized samples are usually inferior to mechanical peeling counterparts. Furthermore, controlling the relative orientations during restacking of monolayers induced uncertainty in the device performance. From a theoretical viewpoint, the contributors to ubiquitous interlayer coupling are still controversial. The weak van der Waals dispersion interaction and the electrostatic coulombic interaction may in some cases be collectively responsible for the interlayer coupling effect. Quantitative characterization of the coupling strength in various two-dimensional materials is still a problem deserving further investigation. A deep understanding of the physics behind interlayer coupling is very useful to develop new strategies for designing new devices. Nonetheless, these new materials and underlying physics offer unprecedented opportunities for both fundamental research and electronics/optoelectronics applications in the future.

**Acknowledgments**

The authors acknowledge funding support from the National Natural Science Foundation of China (under Grant Nos. 61722403 and 11674121), the Recruitment Program of Global Youth Experts in China, the

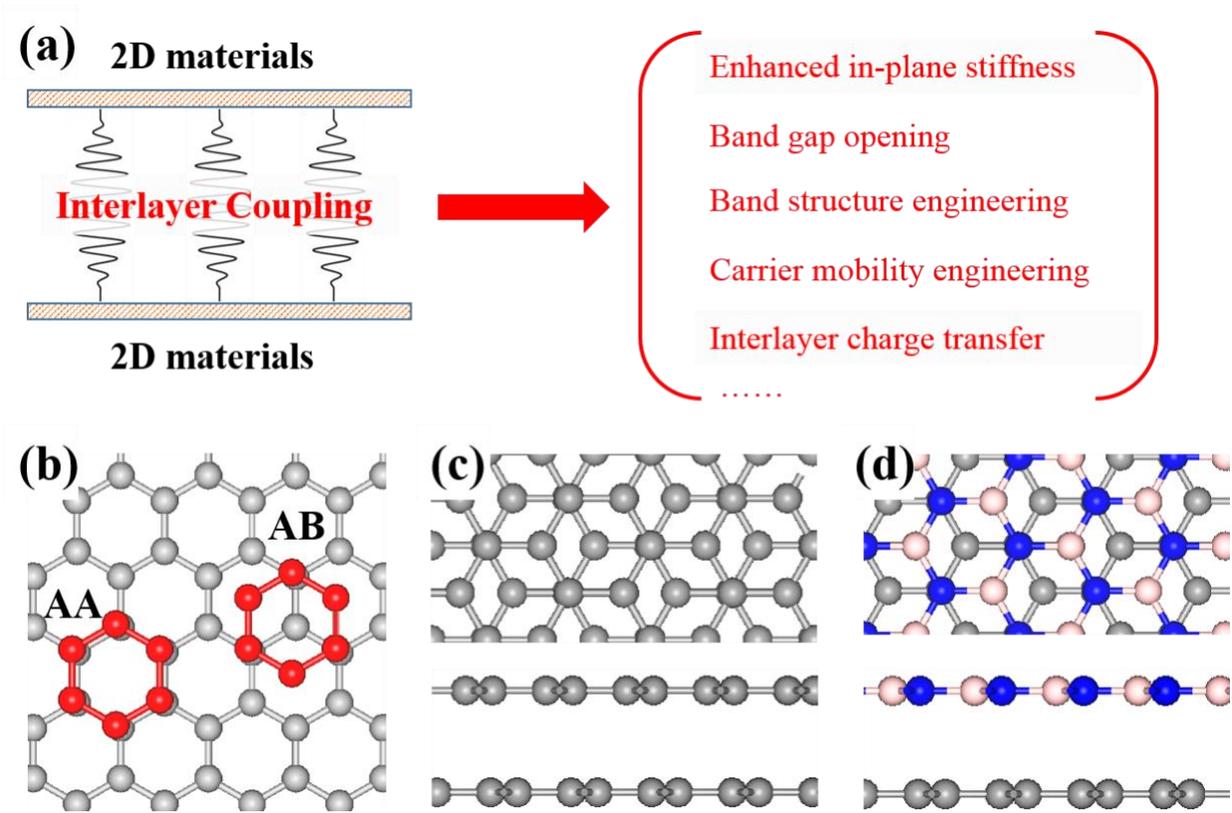

**Fig 1.** A schematic diagram of (a) interlayer coupling and the corresponding effects, (b) AA and AB stacking, and (c) homo and (d) heterostructure models.

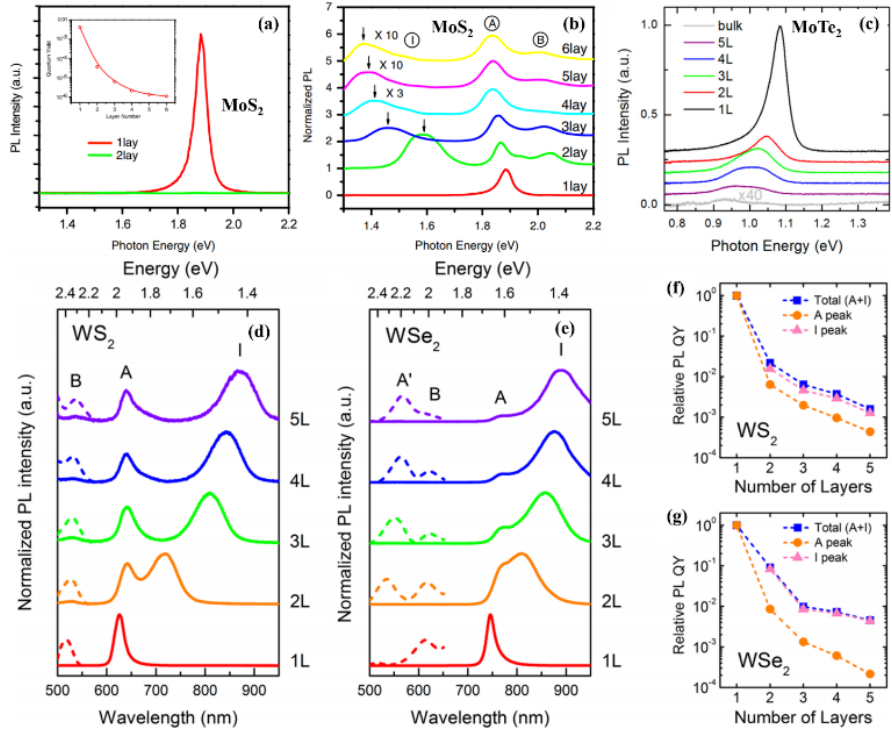

**Fig 2.** (a) PL spectra for mono and bilayer MoS$_2$ samples in the photon energy range from 1.3 to 2.2 eV. Inset: PL QY of thin layers for N = 1–6. (b) PL spectra normalized by the intensity of peak A of thin layers of MoS$_2$ for N = 1–6. Feature I for N = 4–6 is magnified, and the spectra are displaced for clarity. (c) Thickness-dependent photoluminescence (PL) spectra of MoTe$_2$ crystals on SiO2/Si. Bulk data were measured with higher excitation power but normalized in the Fig assuming a linear response. (d, e) Normalized PL spectra of mechanically exfoliated (a) 2H-WS$_2$ and (b) 2H-WSe$_2$ flakes consisting of 1-5 layers. Peak I is indirect gap emission. Weak hot electron peaks A' and B are magnified and are shown as dashed lines for clarity. The total emission intensity becomes significantly weaker with increasing layer number. (f, g) Relative decay in the PL QY with the number of layers for (f) WS$_2$ and (g) WSe$_2$. The plots are shown for the A and I peaks and their sum (A+I). (a, b) Adapted with permission [23].Copyright 2010, APS. (c) Reproduced with permission [102]. Copyright 2014 American Chemical Society. (d-g) Adapted with permission [103]. Copyright 2014 American Chemical Society.

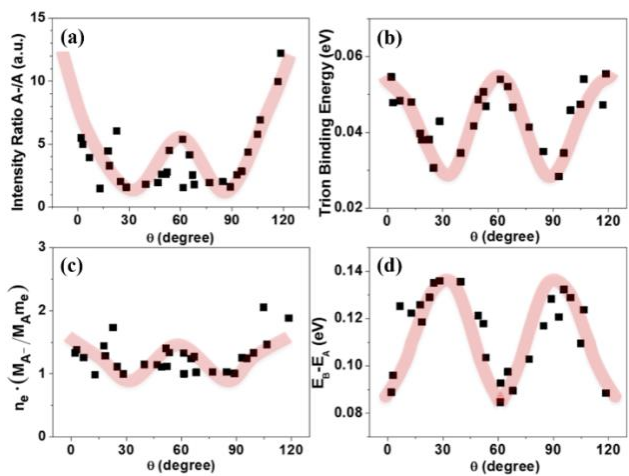

**Fig 3.** Twisted angle dependence of the: (a) A⁻ trion to A exciton PL intensity ratio, (b) trion binding energy $\varepsilon_{A^-}$, (c) $n_e \cdot [M_{A^-}/(M_A \cdot m_e)]$ normalized by the value at $\theta = 30°$, (d) PL peak energy difference between the B and A excitons. The black solid squares are experimental data points, and the pink solid lines show the changing trend of each parameter. Adapted with permission [91]. Copyright 2014 American Chemical Society.

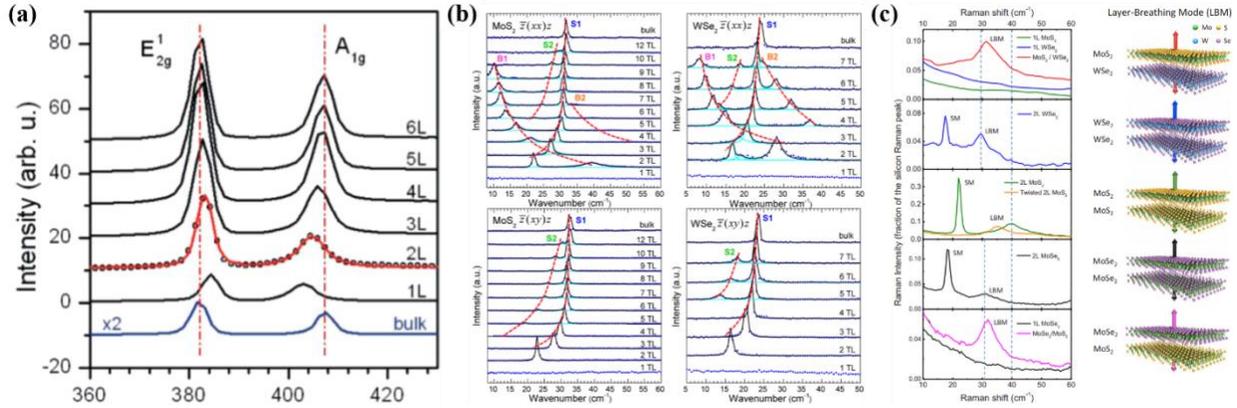

**Fig 4.** (a) Raman spectra of thin (*n*L) and bulk MoS$_2$ films. The solid line for the 2L spectrum is a double Voigt fit through the data (circles for 2L, solid lines for the rest). (b) Low-frequency Raman spectral evolutions as a function of trilayer number in MoS$_2$ and WSe$_2$. Low-frequency Raman spectra of 1−12TL MoS$_2$ (left) measured using (left top) the $\bar{z}$ (xx)z polarization configuration, and (left bottom) the $\bar{z}$ (xy)z polarization configuration. Low-frequency Raman spectra of 1−7TL WSe$_2$ (right) measured under the (right top) $\bar{z}$ (xx)z polarization configuration and (right bottom) $\bar{z}$ (xy)z polarization configuration. The blue dots are experimental data points, while the black solid curves are Lorentzian curves fit to the data. The Rayleigh scattering background was subtracted for all the spectra using a polynomial baseline treatment. (c) Low-frequency Raman spectra (left column) and schematics of the LBM vibrations (right column) for the MoS$_2$/WSe$_2$ hetero bilayer, 1L MoS$_2$, and 1L WSe$_2$; Bernal-stacked 2L WSe$_2$; Bernal-stacked and twisted 2L MoS$_2$; Bernal-stacked 2L MoSe$_2$; and the MoSe$_2$/MoS$_2$ hetero bilayer and 1L MoSe$_2$. The dashed vertical lines highlight the LBM positions of Bernal-stacked 2L WSe$_2$, 2L MoS$_2$, and 2L MoSe$_2$. (a) Adapted with permission [116]. Copyright 2010 American Chemical Society. (b) Adapted with permission [119]. Copyright 2013 American Chemical Society. (c) Adapted with permission [123]. Copyright 2010 APS.

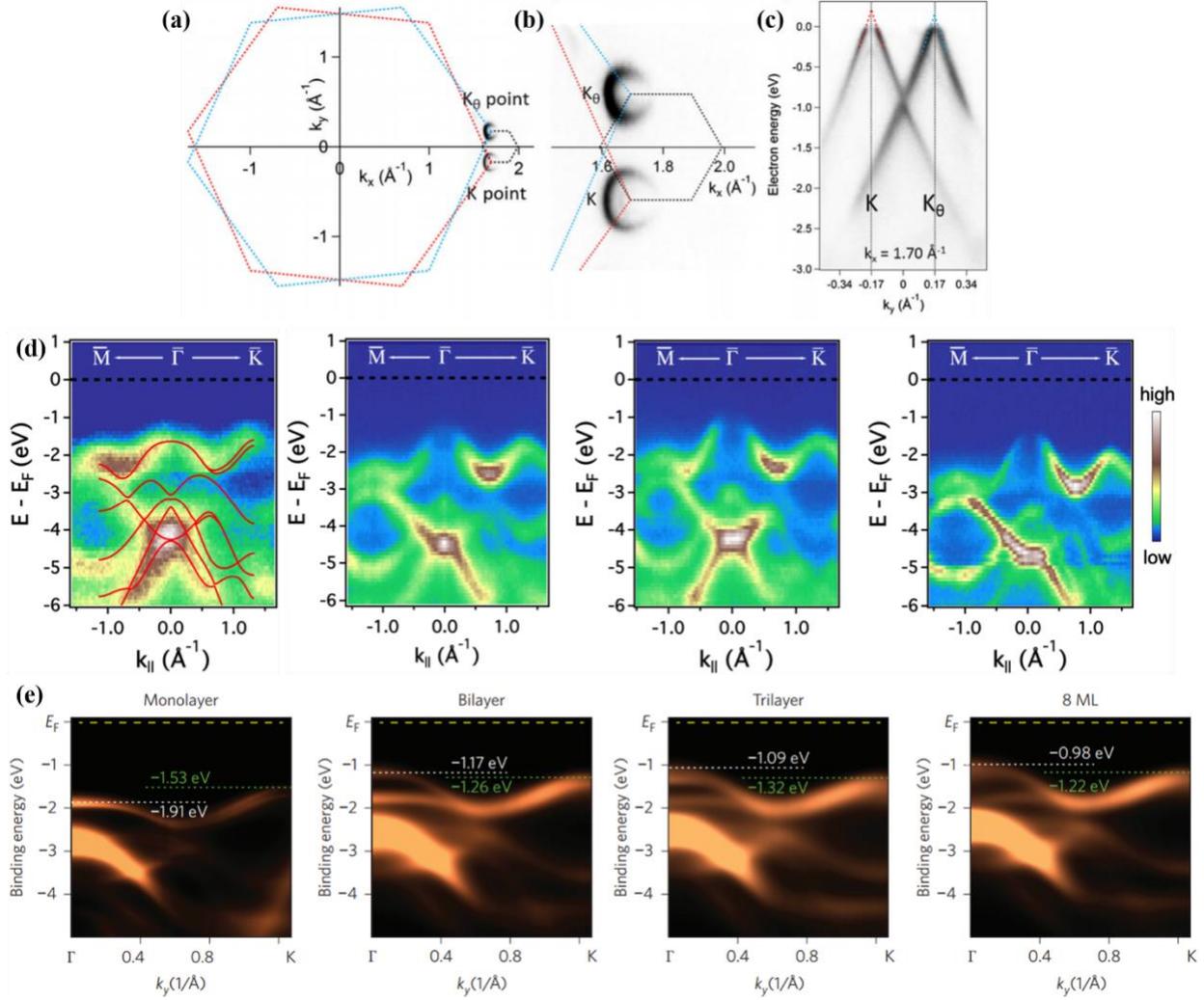

**Fig 5.** (a) The *k* space representation of twisted bilayer graphene with $\theta = 11.6°$, the photoemission intensity contour of the two Dirac cones at the electron energy $E_F$- 0.4 eV, and the primitive BZs of the bottom layer and top layer (red hexagon including the *K* point and blue one including the $K_\theta$ point). Darker shades indicate higher photoemission intensities. The small black hexagon is the Moiré superlattice BZ of the (*p, q*) = (3, 17) commensurate twisted bilayer graphene. (b) Enlarged image of (a) near the two cones. (c) Photoemission spectra intersecting two cones at K and K points. The red and blue dashed lines illustrate the bottom and top layer cones, respectively. (d) ARPES band maps of the exfoliated monolayer, bilayer, trilayer, and bulk $MoS_2$ from left to right, respectively. (e) ARPES spectra of monolayer, bilayer, trilayer and 8 ML $MoSe_2$ thin films along the G to K direction. White and green dotted lines indicate the energy positions of the apices of valence bands at the G and K points, respectively. (a-c) Reproduced with permission [127] Copyright 2012 APS, (d) Reproduced with permission [128]. Copyright 2013 APS. (e) Reproduced with permission [129]. Copyright 2014 Nature Publishing Group.

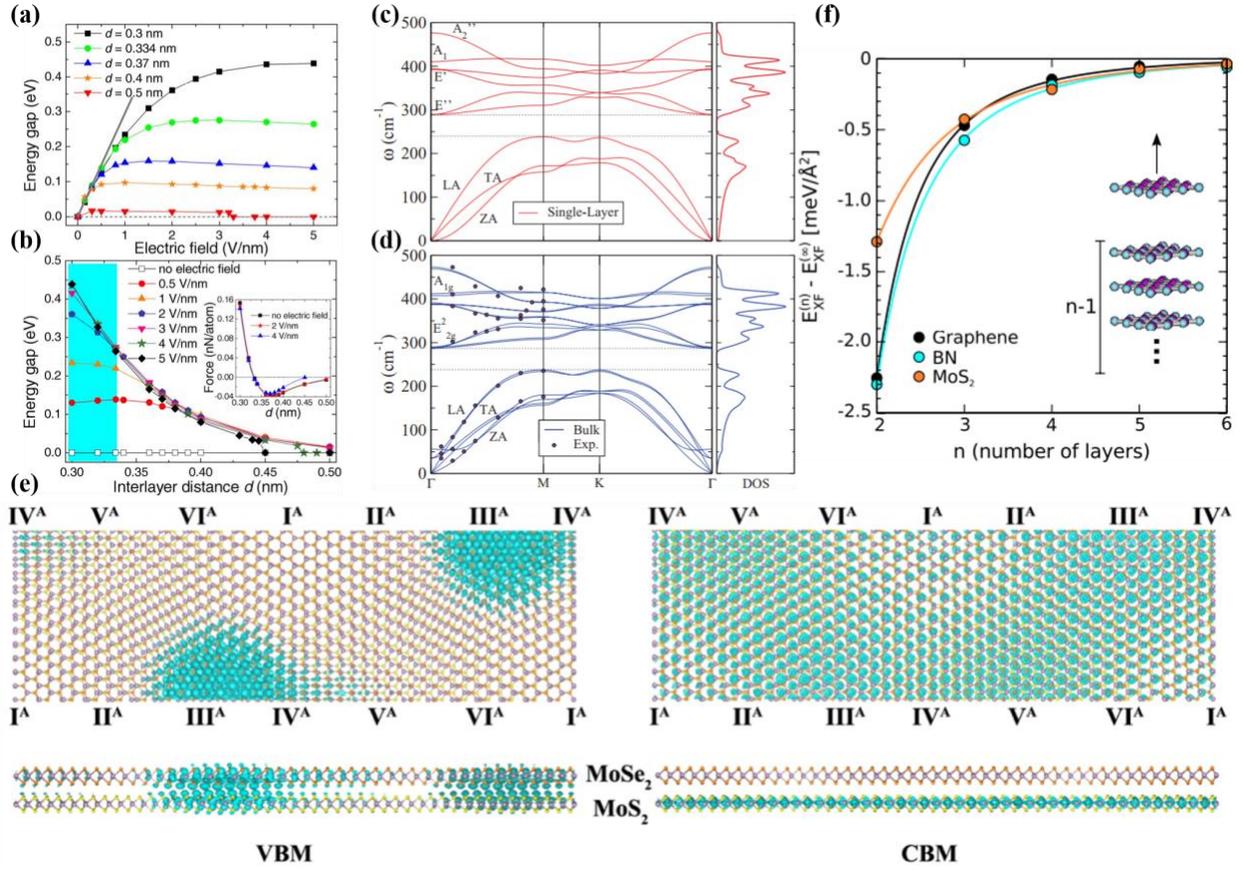

**Fig 6.** (a) Variation in the energy gap with electric field at different interlayer spacings. The gray line on the left side of the curves indicates universal linear scaling of the gap at low fields (below 0.3 V/nm) with a slope of 0.294 eV per V/nm for d ≤ 0.4 nm. (b) Variation of the energy gap with the interlayer spacing in different electric fields. The shaded/clear regions on the left/right side correspond to compression/expansion of the interlayer spacing from equilibrium (d = 0.334 nm). The inset in (b) shows the variation of the force between the two graphene layers. Phonon dispersion curves and density of states of (c) single-layer and (d) bulk $MoS_2$. (e) Top view and side view of the spatial distribution of the VBM, VBM-1 (left), and CBM (right) states for the $MoS_2/MoSe_2$ Moiré pattern. (f) Energy required for exfoliating a single layer from a multilayer structure as a function of the number of layers *n* as shown schematically in the inset figure. (a, b) Adapted with permission [130] Copyright 2008 AIP Publishing, (c, d) Adapted with permission [132] Copyright 2011 APS. (e) Adapted with permission [135] Copyright 2013 American Chemical Society. (f) Adapted with permission [136] Copyright 2012 APS.

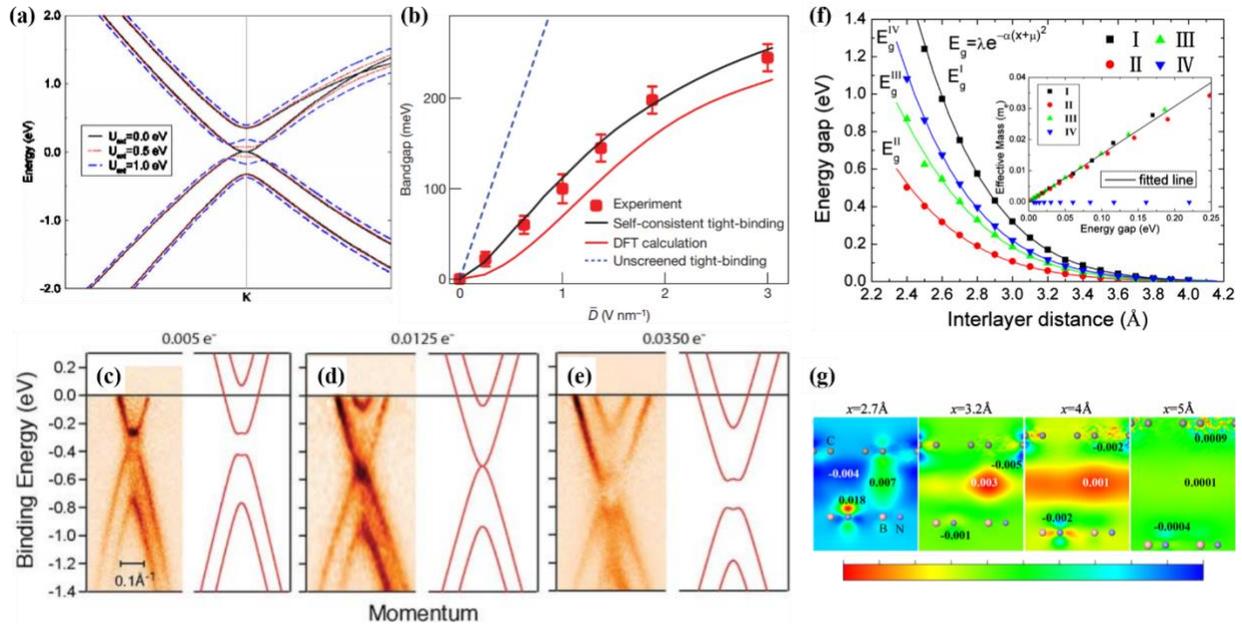

**Fig 7.** (a) Bilayer graphene band structure near the K point for $U_{ext}$ = 0, 0.5, and 1 eV. (b) Experimental and simulation results of the electric-field dependence of the tunable energy band gap in bilayer graphene. Evolution of the gap closing and reopening by changing the doping level via potassium adsorption. Experimental and theoretical bands (solid lines) are shown (c) for as-prepared bilayer graphene and (d and e) with progressive adsorption of potassium. The number of doping electrons per unit cell, estimated from the relative size of the Fermi surface, is indicated at the top of each panel. (f) Variation in the energy gap of the graphene/*h*-BN bilayer as a function of interlayer spacing. The inset shows the variation of the effective mass vs band gap. (g) Contour plots of the charge density difference of AA stacking graphene/*h*-BN bilayer. (a) Reproduced with permission [55] Copyright 2007 APS, (b) Reproduced with permission [137] Copyright 2009 Nature Publishing Group, (c-e) Reproduced with permission [54]. Copyright 2006 AAAS. (f and g) Reproduced with permission [138]. Copyright 2011 AIP Publishing.

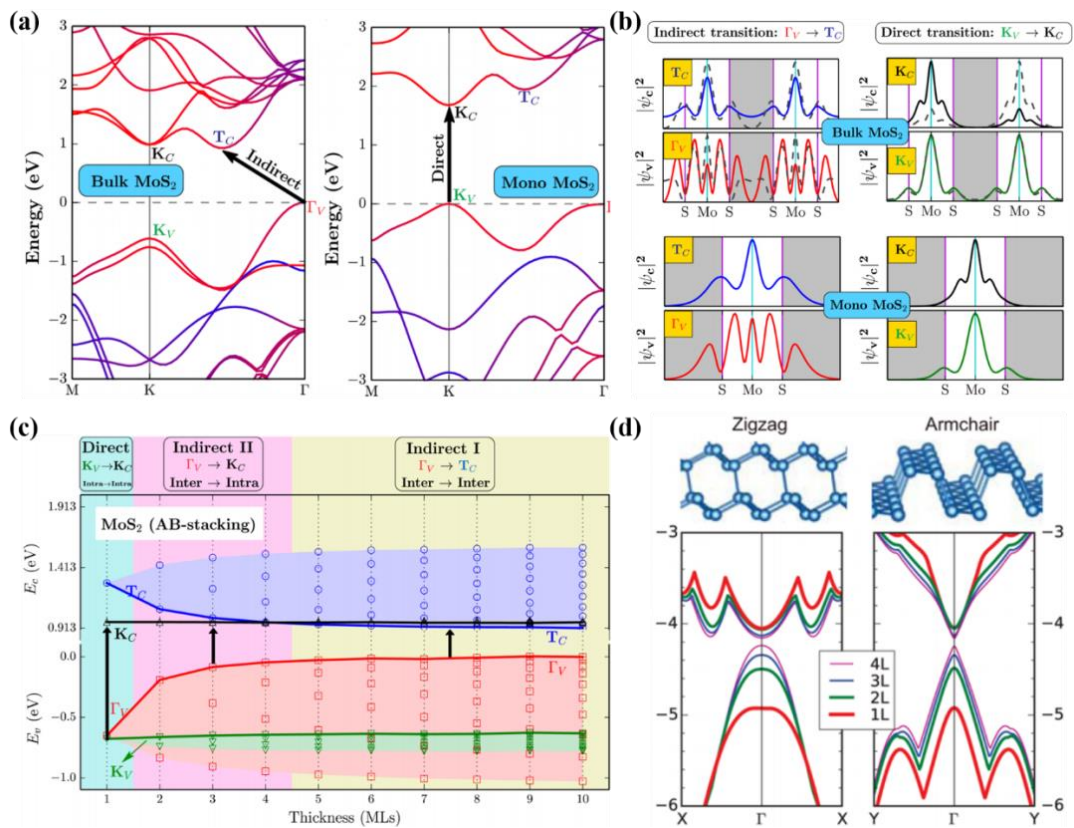

**Fig 8.** (a) Calculated band structure of bulk (left) and the monolayer (right) $MoS_2$. The band structure has been projected onto constituting atomic species with the blue color representing S and red representing Mo. The band-edge states relevant to the optical transition ($K_V$, $K_C$, $\Gamma_V$, and $T_C$) are labeled, and the band gap transitions are indicated. (b) Planar-averaged squared magnitude of the wavefunctions for the relevant band-edge states bulk (upper panels) and monolayer (lower panels) $MoS_2$ plotted along the direction perpendicular to the layers. The positions of the Mo and S atoms are marked, and the interstitial region outside the sandwich S−Mo−S layers is shown in gray. In the upper panels, the second band-edge state is shown with the dark gray dash line. (c) Evolution of various band-edge states of $MoS_2$ going from multiple-layered films (with the AB-type stacking) to the monolayer. The energy levels corresponding to different band edges are shown in different colors: $\Gamma_V$ (red), $K_V$ (green), $T_C$ (blue), and $K_C$ (Black). For each state, the span of energy levels is shaded in a lighter color. The valence band maximum of the $\Gamma_V$ state at 10 MLs is set to zero. The first y-axis tick of the upper panel (conduction bands) refers to the band gap value at 10 MLs. The band gap transitions show three distinct regimes (see text for more detailed description): Indirect I ($\Gamma_V \rightarrow T_C$, in yellow background), Indirect II ($\Gamma_V \rightarrow K_C$, in pink), and Direct ($K_V \rightarrow K_C$, in cyan). (d) The evolution of the band-edge states along the zigzag and armchair direction as a function of the number of layers. (a-d) Adapted with permission [146, 148]. Copyright 2015 American Chemical Society.

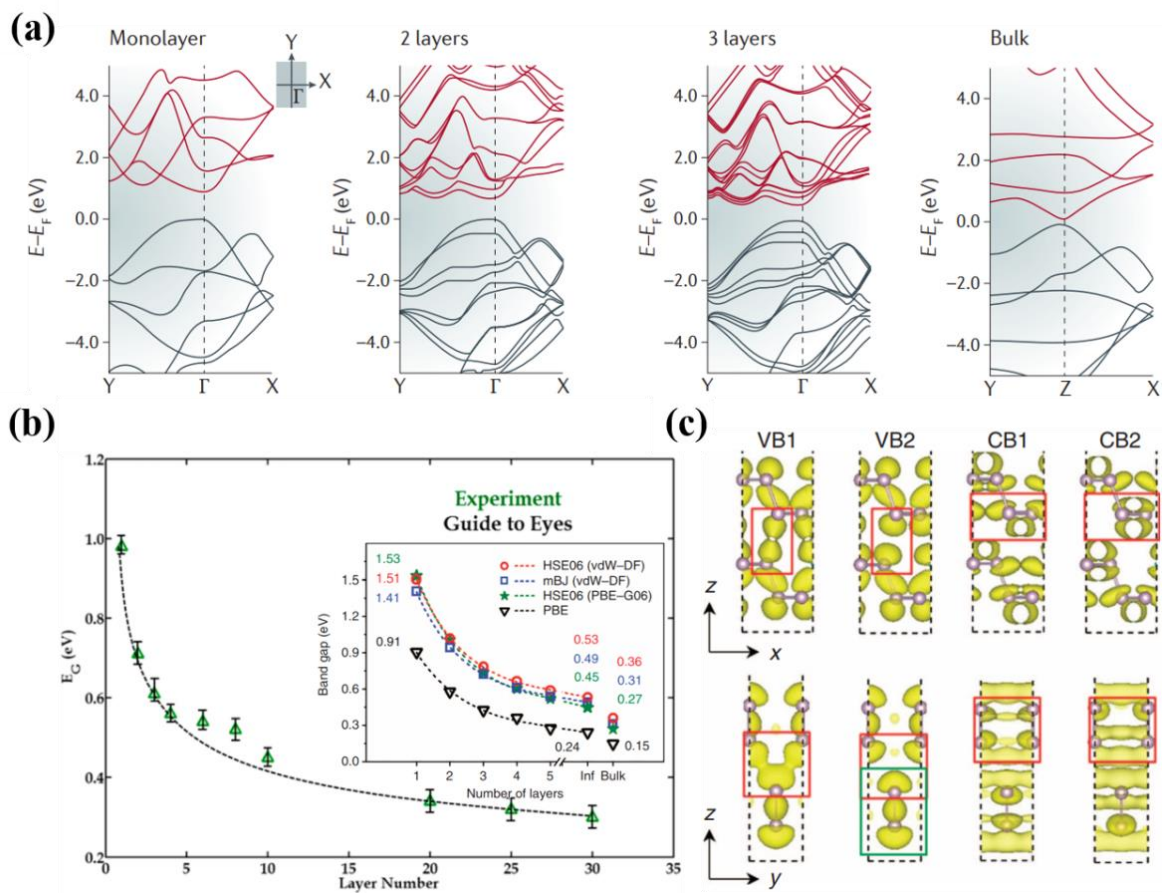

**Fig 9.** (a) Band structures obtained by density functional theory for mono-, bi- and trilayer phosphorene and for bulk black phosphorus (from left to right); Γ = (0, 0) denotes the center of the 2D Brillouin zone. (b) Evolution of the direct band gaps as a function of sample thickness. Inset is the theoretical calculation results, and the functionals used for structural optimization are shown in parentheses. (c) Spatial structure of wavefunctions for the four marked states illustrated in the xz and yz planes using an isosurface of 0.0025 e/Å$^3$. (a) Reproduced with permission [151] Copyright 2016 Nature Publishing Group. (b, c) Reproduced with permission [153, 152]. Copyright 2014 Nature Publishing Group and American Chemical Society.

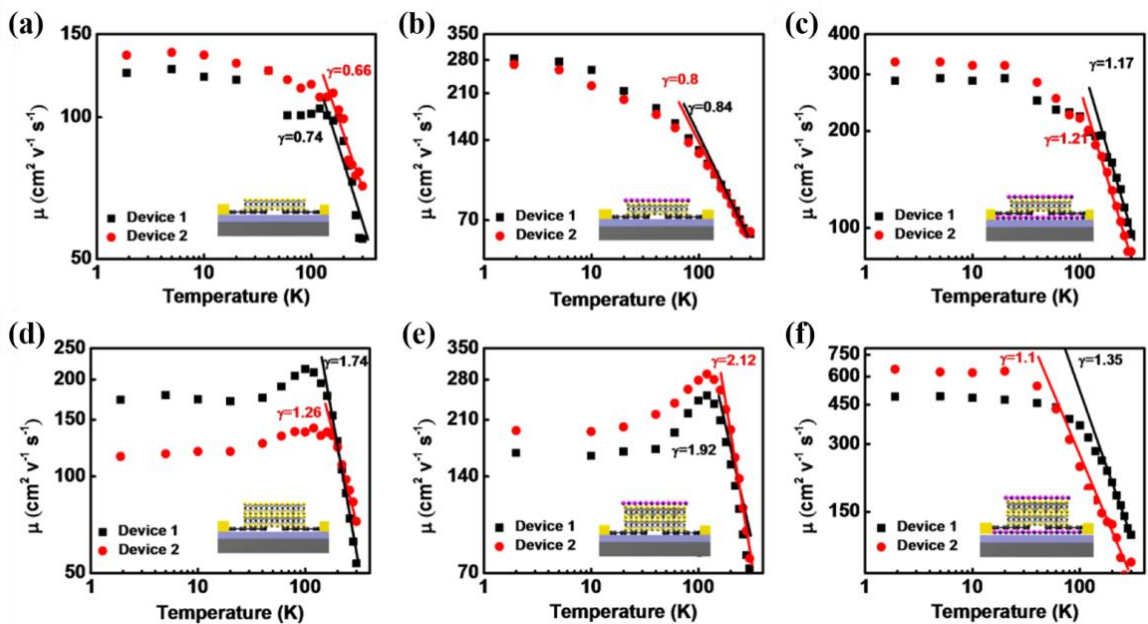

**Fig 10.** Mobility engineering by BN encapsulation. (a,d) Extrinsic field-effect mobility of monolayer and multilayer $MoS_2$ devices as a function of temperature (300−1.9 K). Linear fitting is used in the phonon control region (100−300 K) to extract γ. (b,e) Extrinsic field-effect mobility of monolayer and multilayer devices with top BN encapsulation. (c,f) Extrinsic field-effect mobility of monolayer and multilayer $MoS_2$ devices with bottom and top BN encapsulation, forming a sandwich structure. The figure is adapted with permission [93]. Copyright 2015 American Chemical Society.

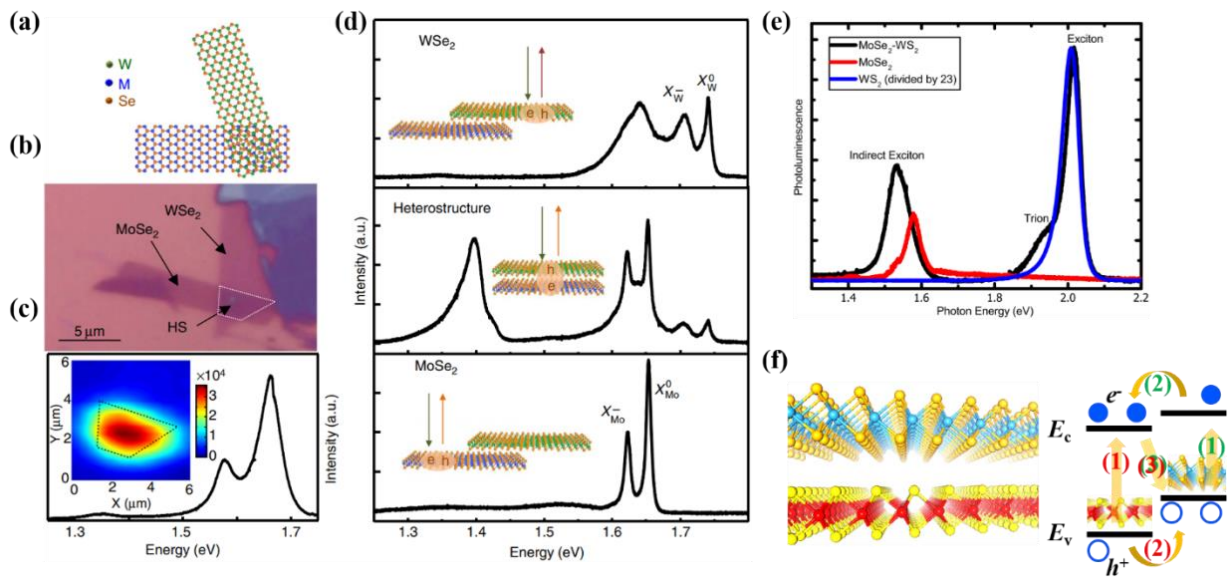

**Fig 11.** (a) Cartoon depiction of a MoSe$_2$–WSe$_2$ vdWHS. (b) Microscope image of a MoSe$_2$–WSe$_2$ vdWHS with a white dashed line outlining the vdWHS region. (c) Room temperature photoluminescence of the vdWHS under 20 mW laser excitation at 2.33 eV. Inset: spatial map of integrated PL intensity from the low-energy peak (1.273–1.400 eV), which is only appreciable in the vdWHS area, outlined by the dashed black line. (d) Photoluminescence of individual monolayers and the vdWHS at 20 K under 20 mW excitation at 1.88 eV (plotted on the same scale). (e) Photoluminescence spectra of the WS$_2$ monolayer (blue, divided by 23), MoSe$_2$ monolayer (red), and MoSe$_2$-WS$_2$ vdWHS (black) obtained under the same conditions. (f) Diagram of interlayer excitons of type II band alignment in TMDCs vdWHS. (a-d) Reproduced with permission [158] Copyright 2015 Nature Publishing Group. (e) Reproduced with permission [159]. Copyright 2015 American Chemical Society.

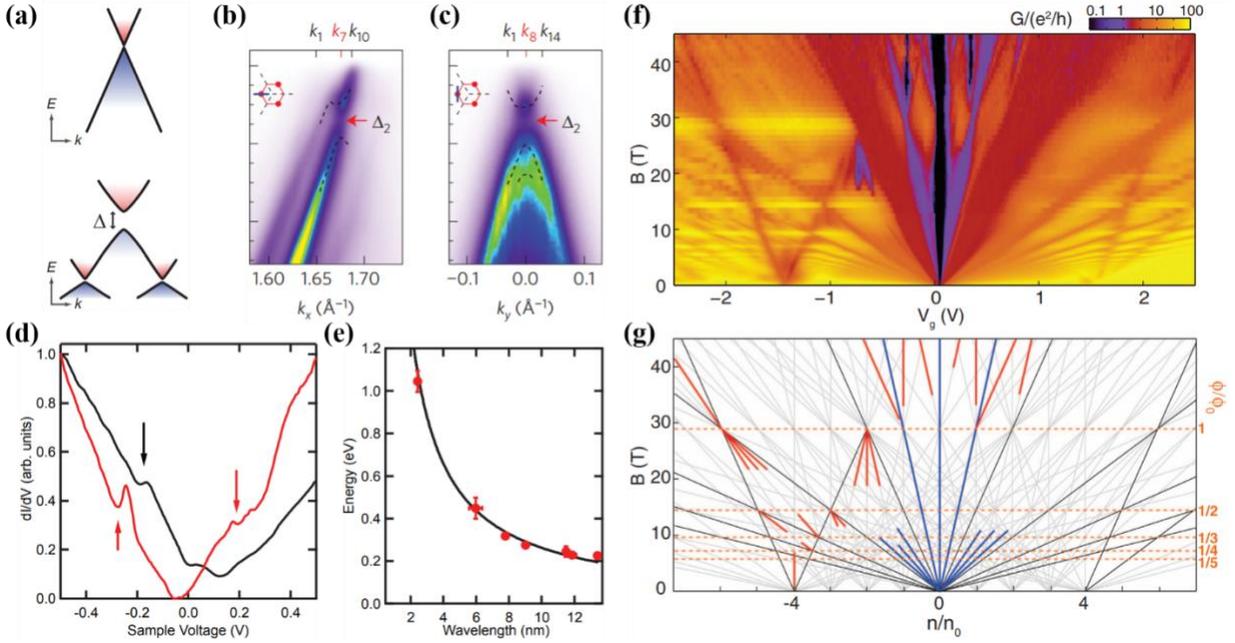

**Fig 12.** (a) Schematics of the band structure of graphene (upper panel) and graphene/$h$-BN vdWHS (under panel). (b, c) ARPES data through an SDC along different directions. The red arrows point to the positions at SDC. (d) Experimental d$I$/d$V$ curves for two different Moiré wavelengths, 9.0 nm (black) and 13.4 nm (red). (e) Energy difference between the SDCs and the FDC diminishes with increased wavelength. (f) Two-terminal magnetoconductance of sample up to 45 T. (g) Energy gaps in the Hofstadter spectrum; gaps intersect at $\varphi/\varphi_0 = 1/q$, where q is an integer and $\varphi/\varphi_0$ at 29 T. (a, f, g) Reprinted with permission from [142] Copyright 2013 AAAS. (b, c) Reprinted with permission from [164] Copyright 2016 Nature Publishing Group. (d, e) Reprinted with permission from [100] Copyright 2012 Nature Publishing Group.

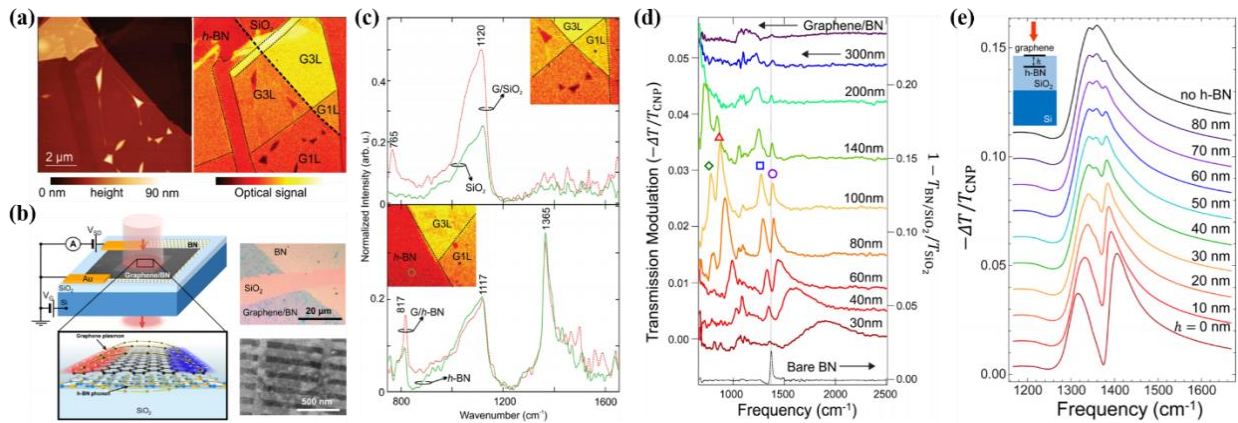

**Fig 13.** (a) AFM topography overview image of a G/h-BN vdWHS lying on a SiO$_2$ substrate. (b) Schematic of G/h-BN nanoresonators, (c) synchrotron infrared nanospectroscopy spectra of G/SiO$_2$ and G/h-BN, as marked in the inset. (d) Normalized transmission spectra of graphene nanoresonators. The dotted vertical line indicates this peak position as a reference for the other spectra. (e) Calculated transmission spectra for 80 nm G/SiO$_2$/h-BN/SiO$_2$ ribbons with different SiO$_2$ layer thicknesses. (a and c) Adapted with permission [182]. Copyright 2015 Royal Society of Chemistry. (b, d, e) Adapted with permission [181]. Copyright 2014 American Chemical Society.

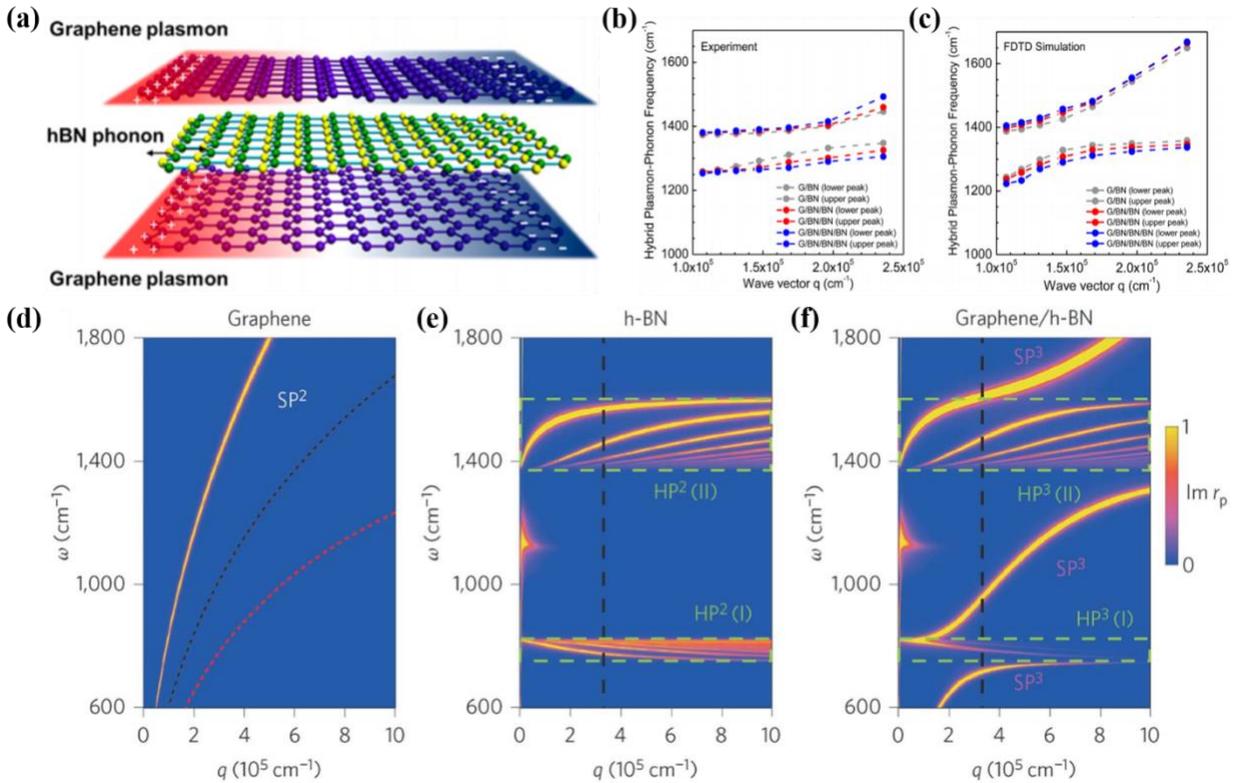

**Fig 14.** (a) Schematic of G/h-BN vdWHS. (b,c) Experimental and simulated plasmon−phonon polariton dispersions of vdWHS with a different number of h-BN layers. (d-f) Calculated dispersion of $SP^2$, $HP^2$ and $HP^3$ in free-standing graphene or h-BN and in G/h-BN vdWHS. (a-c) Reprinted with permission from [183]. Copyright 2014 American Chemical Society. (d-f) Reprinted with permission from [178]. Copyright 2015 Nature Publishing Group.